\documentclass[prl,aps,twocolumn,showpacs,epsf]{revtex4}
\usepackage{graphicx}
\usepackage{epsfig}
\usepackage{epsf}
\usepackage{amssymb}
\usepackage{epstopdf}
\usepackage{amsmath}
\usepackage{amstext}
\usepackage{latexsym}
\usepackage[matrix,frame,arrow]{xypic}

\newcommand{\ket}[1]{|#1\rangle}

\begin{document}

\title{Repeat-Until-Success Generation of Symmetric States without External Control}
\author{ Qing Chen$^{1}$}
\author {Jianhua Cheng$^1$}
\author{ Ke-Lin Wang$^1$}
\author {Jiangfeng Du$^{2,3}$}

\address{
$^{1}$Department of Modern Physics, University of Science and Technology of China, Hefei 230026, PR China\\
$^{2}$Hefei National Laboratory for Physical Sciences at Microscale and Department of Modern Physics,
University of Science and Technology of China, Hefei, Anhui 230026, PR China\\
$^{3}$Fachbereich Physik, Universitaet Dortmund, 44221 Dortmund, Germany
}
\date{\today}

\begin{abstract}
In the present paper, we propose a ``repeat-until-success" scheme induced by single particle measurement to generate arbitrary symmetric
states based on spin network. This protocol requires no modulated controls during the whole process and it provides a persistent approach
towards the desired symmetric state. As a special case, we demonstrate that $W$ state can be created with unit probability within this
framework.
\end{abstract}
\pacs{03.67.Hk, 03.67.-a} \maketitle

%%%%%%%%%%%%%%%%%%%%%%%%%%%%%Main Body%%%%%%%%%%%%%%%%%%%%%%%%%%%%%%%%%%%%%
Ever since the birth of the remarkable Einstein-Podolsky-Rosen (EPR) paradox \cite {EPR}, quantum entanglement, a unique quantum feature,
has attracted much concern in modern physics. This interesting feature reveals many novel phenomena that can
only be interpreted within the field of quantum theory, therefore it provides a natural resource to conduct fundamental test of quantum
mechanics. In recent years, it plays a key role in the emerging technologies of quantum information processing, such as quantum
teleportation, quantum computation and quantum cryptography \cite{applications}.

The symmetric states, a group of highly entangled states which are invariant under arbitrary permutation, have drawn increasing attentions
from some relevant fields. The unique entanglement features of this kind of states are revealed through the study of concurrence \cite{Wang}
and geometric measure \cite{Geometric}. Moreover, symmetric states play an important role in reaching optimal approximate quantum cloning
\cite{Gisin, Chen} and precise spectroscopy measurement \cite{Huelga}.
They also relate to some models in the fields such as quantum phase transition \cite{Vidal} and high temperature superconductor
\cite{superconduct}.
Here the symmetric states are noted as $|S(M,k)\rangle=\frac{1}{\sqrt{C_M^k}}(\hat{P}|\underbrace{000}_k...\underbrace{11}_{M-k} \rangle)$,
where $\hat{P}$ is the total permutation operator, $M$ stands for the total qubit number while there are $k$ qubits in
state $|0\rangle$ and $M-k$ qubits in state $|1\rangle$ ($|0\rangle$ and $|1\rangle$ corresponds to the eigenstates of $\sigma_z$ with
positive and negative eigenvalue respectively). $C_M^k$ denotes the combination number. Therefore
the preparation of a specific $M$ particle symmetric state is a crucial step to experimentally implement certain tasks. Recent progress
has suggested several ways to produce such symmetric states via different systems \cite {Kaye,Unanyan}. These proposals do not
only require interaction between qubits but also elaborate external control of this interaction.

Recently, it has been shown that there are routes to conduct certain computation tasks through quantum evolution without external
controls upon a properly designed spin network \cite{Chen, Bose, Perfecttransfer, Other, SpinCompute, Clone}.
Bose initialed the state transfer implemented in a $1$D spin chains with the conventional Heisenberg Hamiltonian \cite{Bose} and was
later demonstrated that perfect state transfer is available within this framwork \cite{Perfecttransfer}.
While star-like spin network has been used to implement approximate quantum cloning by De Chiara \emph{et al} \cite{Clone},
Chen \emph{et al} further prove that optimal phase covariant clone (PCC) is available in the same physical system if only the
initial state of supplementary qubits were prepared as a specific symmetric state \cite{Chen}.

This framework has also been employed to create a large group of entangled states including the cluster states \cite{Briegel}.
In this Letter, we hereby propose a single particle measurements induced construction of the symmetric states with a properly
designed spin network. In this scheme, with the help of single particle measurements performed upon
the supplementary system, the state of the target system is supposed to collapse to a specific symmetric state with a certain probability.
More important, our scheme further provides a ``repeat-until-success" mode in which the whole preparation process can be iterated
until we reach the required symmetric state. And the system does not even need to be recovered to the initial state before we begin
the next round's attempt in case the first round attempt fails. The detailed scheme will be addressed in the sequel.

The spin network utilized in this paper is shown as Figure $1$b, and the corresponding Hamiltonian is defined as
\begin{eqnarray}
H &=& \frac{\cal{J}}{2}\sum_{i=1}^{i=N}\sum_{j=1}^{j=M}(\sigma_{1i}^x\sigma_{2j}^x+\sigma_{1i}^y\sigma_{2j}^y
+\lambda \sigma_{1i}^z\sigma_{2j}^z) \nonumber\\
&+&\frac{B}{2}\sum_{i=1}^{i=N}\sum_{j=1}^{j=M}(\sigma_{1i}^z+\sigma_{2j}^z),
\end{eqnarray}
where $\sigma_{1i}^{x, y, z}$ denotes the Pauli matrices of the $i$th spin of supplementary system while $\sigma_{2j}^{x, y, z}$ denotes
the $j$th spin of target system.
The supplementary system is shown shown on the left side in Figure $1$b while the target system on the right side. ${\cal{J}}$ stands for
the coupling strength and $B$ is the externally applied magnetic field.
$\lambda$ is the anisotropy parameter which stands for the coupling strength of $z$ direction. When $\lambda=0$, the above Hamiltonian
reduces to $XX$ model while when $\lambda=1$, it turns out to be the standard Heisenberg Hamiltonian.
Obviously, this Hamiltonian can be described as the interaction between a spin-$\frac{M}{2}$ and a spin-$\frac{N}{2}$ if we take
 $\vec{S_1}$ and $\vec{S_2}$ denoting the total spin of $N$ supplementary particles and $M$ target particles respectively (
$N\geq\frac{M}{2}$).
Specifically, $S_{1x, y, z}=\sum_{i=1}^{i=N}\sigma_{1i}^{x, y, z}, S_{2x, y, z}=\sum_{i=1}^{i=M}\sigma_{2i}^{x, y, z}$. The Hamiltonian
can be rewritten as $H=2{\cal{J}}(S_{1x}S_{2x}+S_{1y}S_{2y}+\lambda S_{1z}S_{2z})+ B(S_{1z}+S_{2z})$.
The total angular momentum is defined as $\vec{S}=\vec{S_1}+\vec{S_2}$, and $S, S_1, S_2$ are the quantum numbers associate to the
corresponding operators. Note this Hamiltonian also preserves the $z$ opponent of the total angular momentum.
\begin{figure}
\epsfig{file=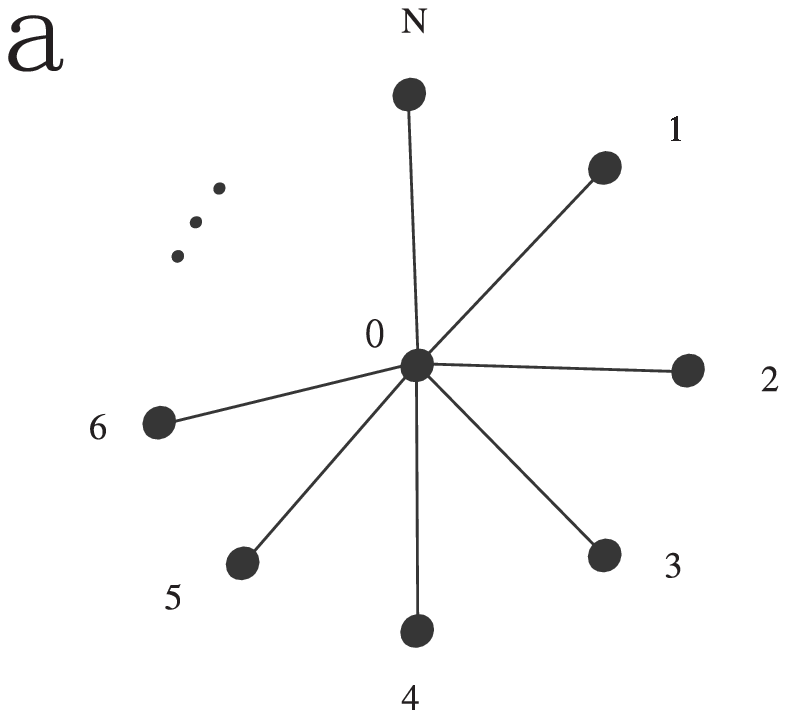,width=0.25\textwidth,height=4cm}
\epsfig{file=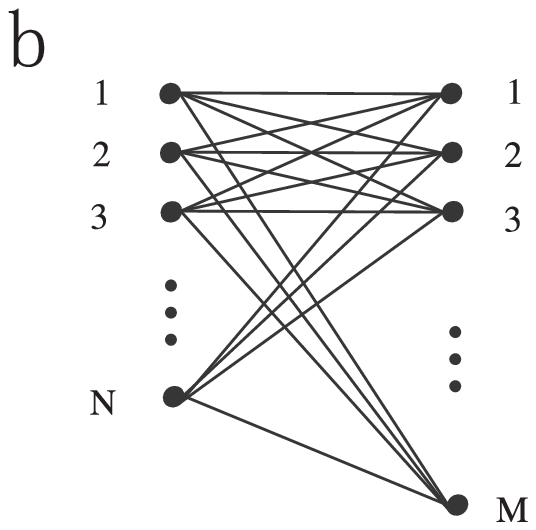,width=0.2\textwidth,height=4cm}
\caption{
(a) A star-like spin network in which the center spin interacts with all the outside spins.
(b) The left side spins form the supplementary system, and spins on the right side form the
target system. Each spin from the supplementary system interacts with all the spins in the target system.
When $N=M$, the above spin structure forms a mirror spin network. }
\end{figure}

The initial state of the system is prepared as
\begin{equation}
\ket{\psi_i}=\ket{11...1}_N\ket{00...0}_M,
\end{equation}
which can be rewritten as $\ket{\psi_i}=\ket{\frac{N}{2}, -\frac{N}{2}}\ket{\frac{M}{2}, \frac{M}{2}}$ in the terms of non-coupled basis
$\ket{S_1, S_{1z}}\ket{S_2, S_{2z}}$. Note that in this basis, symmetric states can be written as
$\ket{S(M, k)}=\ket{\frac{M}{2}, k-\frac{M}{2}}$ ($0\leq k\leq M$).
The initial state is a product state which makes the experimental implementation easily accessible.
A careful analysis of the Hamiltonian indicates that the state of the system after evolution will navigate in a subspace spanned by
$N+1$ base vectors $\{\ket{\frac{N}{2}, -\frac{N}{2}}|\frac{M}{2}, \frac{M}{2}\rangle, ... ,
\ket{\frac{N}{2}, \frac{N}{2}}|\frac{M}{2}, \frac{M-2N}{2}\rangle   \}$. Obvious observation reveals that the state of the target system
navigates in the subspace of $\{\ket{S(M,k)}, k=0,1..., M\}$.
As the number of particles increases, it is unlikely that the total state of the system evolve to a certain
$\ket{\frac{N}{2}, (M-k)-\frac{N}{2}}|\frac{M}{2}, k-\frac{M}{2}\rangle$ from the initial product state by simply controlling the limited
parameters of the Hamiltonian.
However, by performing single particle measurements upon the supplementary system on $z$ direction, the state of the target system is to
collapse to a specific symmetric state.
Moreover, according to the conservation of $z$ opponent of the total angular momentum, the result of the measurements could tell the
exact symmetric state to which the target system collapsed. For instance, if there appears to be $(M-k)$ particles spin-up after the
measurements, the state of the target system is undoubted to be the required symmetric state $S(M, k)$. Since the collapsing of quantum
states is a random case, one could only reach the required state with a certain probability.
When noticed that the state of target system remains navigating in the $\{\ket{S(M,k)}, k=0,1..., M\}$ subspace no matter
how many times the supplementary system been measured, we can further propose a persistent
scheme to ensure the protocol's success, namely the ``repeat-until-success" scheme. Under this scheme, what one has to do is to keep
performing $z$ direction single particle measurements upon assistant particles, and identify the state of the target system. If the
required state has been generated successfully, the process ends. Otherwise, let the whole system evolve continually, and choose
a proper later time (so that a relative greater success probability is expected) to begin the next round measurements. This process can
be iterated until the required symmetric state is successfully obtained. Therefore the ``repeat-until-success" scheme provides a promising
strategy to create arbitrary symmetric states based on spin networks.

In the following, for simplicity, we will discuss the maximum success probability in a mirror spin network where both supplementary
system and the target system have $N$ particles (here $N$ is assumed to be an even number). According to our calculation,
the maximum success probability approaching to the specific symmetric state $\ket{S(N, \frac{N}{2})}$ within this special spin
architecture can be achieved.
The corresponding Hamiltonian can be expressed as ( the coupling strength {${\cal{J}}$ is set as $1$)
\begin{equation}
H=2\vec{S_{1}}\cdot\vec{S_{2}}+B(S_{1z}+S_{2z}),
\end{equation}
where $\vec{S_1}=\frac{1}{2}\sum_{i=1}^{N}\vec{\sigma_{1i}}, \vec{S_2}=\frac{1}{2}\sum_{j=1}^{N}\vec{\sigma_{2j}}$.
The eigenstates of this
Hamiltonian can be written as $\ket{S,S_1,S_2,S_z}$ in terms of the corresponding quantum numbers with the associate eigenvalue
$E=S(S+1)-S_1(S_1+1)-S_2(S_2+1)$. The initial state of the whole system is prepared to the product state
$\ket{\frac{N}{2}, -\frac{N}{2}}\ket{\frac{N}{2}, \frac{N}{2}}$, which is expressed in terms of $\ket{S_1, S_{1z}}\ket{S_2, S_{2z}}$.
Straightforward calculation provides the state of the system after a period of evolution
\begin{eqnarray}
\ket{\psi_t}
=\sum_{{\small m=-\frac{N}{2}}}^{\frac{N}{2}} \left(\sum_{S=0}^{N}\exp(-iE_S\;t)P_{S,\;m}\right) \ket{\frac{N}{2}, m}\ket{\frac{N}{2}, -m},\nonumber\\
\end{eqnarray}
where $E_S=S(S+1)-\frac{N}{2}(N+2)$,
\begin{eqnarray}
P_{S,m} &=& \langle S, \frac{N}{2}, \frac{N}{2}, 0|\frac{N}{2}, -\frac{N}{2}\rangle
\ket{\frac{N}{2}, \frac{N}{2}}\nonumber\\
&&\langle \frac{N}{2}, m|\langle \frac{N}{2}, -m|S, \frac{N}{2}, \frac{N}{2}, 0\rangle.
\end{eqnarray}
Therefore, if $z$ direction single particle measurements be performed upon the supplementary system, the probability that the state of
target system collapsed to $\ket{\frac{N}{2},0}$ is supposed to be
\begin{eqnarray}
P(t)&=&|\sum_{S=0}^{N}\exp(-iE_S\;t)P_{S,\;0}|^2 \nonumber \\
&\leq& \sum_{S=0}^{N}|P_{S,\;0}|^2.
\end{eqnarray}
With the help of C-G (Clebsch-Gordan) form, one could work out the explicit expression of $P_{S,\;0}$
\begin{equation}
P_{S,\;0}=
\frac{(-1)^{\frac{N+S}{2}}2^{-N-2}(2S+1)N!\;\Gamma(\frac{S+1}{2})(1+\cos[\pi S])}
{\Gamma(\frac{N-S+2}{2}) \Gamma(\frac{S}{2}+1)\Gamma(\frac{N+S+3}{2})} \nonumber.
\end{equation}
After detailed calculation, it is proved that the success probability $P(t)$ can saturate its optimal value when $t=\pi$. And the maximum
success probability is
\begin{eqnarray}
P_{m} &= &\left\{\Gamma(\frac{N+1}{2}) \right.\nonumber\\
&&\left. \left[\frac{\sqrt{2\pi}}{\Gamma(-\frac{1}{4})\Gamma(\frac{2N+5}{4})} +
 \frac{N\;{_2}F_1(\frac{3}{2}, 1-\frac{N}{2}, \frac{N+5}{2}, -1)}{2\;\Gamma(\frac{N+5}{2})}\right]\right\}^2, \nonumber\\
\end{eqnarray}
where ${_2}F_1(a, b, c, z)$ stands for regularized hypergeometric function and $\Gamma(x)$ stands
for Gamma function.

Figure 2 describes the optimal success probability goes with particle number $N$. The figure indicates that, as particle
number increases, the expected optimal success probability tends to decrease. nevertheless, the variance ratio slows down when $N>20$.
Actually, the maximum success probability still holds above $0.1$ when spin number is $40$.
Here we just pick up an example with the standard Heisenberg Hamiltonian.
In fact, arbitray $XXZ$ model can be utilized to create symmetric states under the same framework.
In principle, once the initial parameters are given ( the spin network, initial state of the system etc.),
one could further work out a series of timetable which points out the exact moment measurements should be conducted upon supplementary
system. Therefore, within the ``repeat-until-success" scheme, such a timetable ensures that maximum success probability is
reached in each round's attempt to achieve the required symmetric state.
\begin{figure}
\epsfig{file=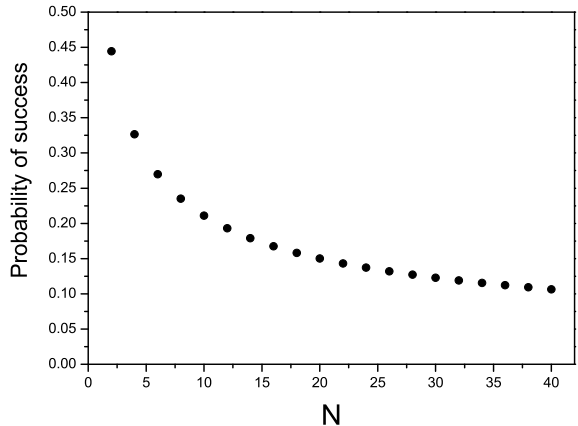, width=0.5\textwidth, height=7cm}
\caption{ The optimal success probability after the first round of measurements performed upon supplementary system in the approach
towards $\ket{S(N,N/2)}$ based on the mirror spin network ($N$ is an even number).}
\end{figure}

Finally, we point out that the $W$ state, a special case of symmetric states, can be generated through dynamic evolution with unit
probability based on a star like spin network shown as Figure $1$a. Note that similar scheme has been proposed in qubit-cavity system
\cite{Alexandra}. The Hamiltonian involved here is the conventional $XXZ$ model defined as
\begin{equation}
H=\frac{\cal{J}}{2}\sum_{i=1}^{M}(\sigma_0^x\sigma_{i}^x+\sigma_0^y\sigma_{i}^y+
\lambda\;\sigma_0^z\sigma_{i}^z)+\frac{B}{2}\sum_{i=0}^{M}\sigma_{i}^z.
\end{equation}

To generate $M$-particle $W$ state, the $XX$ model is good enough. The initial state of the system is prepared as
$|\psi(0)\rangle=\ket{1}\ket{00...0}=|1\rangle|M/2, M/2\rangle$, a product state which can be implemented easily in practical experiments.
To achieve $M$-particle $W$ state $S(M,M-1)$, one possible way is to prepare it upon the outside spins.
For the very special case where $\lambda=0,B=0$, the eigenstates and corresponding eigenvalues are reduced to
\begin{eqnarray}
|\psi\rangle_{j,m-\frac{1}{2}}^{\pm}&=&|0\rangle |j,m-1\rangle \pm|1\rangle |j,m\rangle, \nonumber\\
E_{j,m-\frac{1}{2}}^{\pm}&=&\sqrt{(j+m)(j-m+1)}.
\end{eqnarray}
Hence, the time dependent state of the system is
\begin{equation}
\ket{\psi_t}=\cos \sqrt{M}t |1\rangle|\frac{M}{2}, \frac{M}{2}\rangle
+i\sin \sqrt{M}t \ket{0}\ket{\frac{M}{2},\frac{M}{2}-1}.
\end{equation}
Straightforward calculation yields that the outcome state of the outside $M$ spins is the desired $W$ state after
the evolution if evolution time satisfies $t=\frac{\pi}{2\sqrt{M}}$.
Obviously, the center spin is wasted. Is it possible that all spins are made use of in the generation of $W$ states? The answer is
positive. The $M+1$-particle $W$ state can be rewritten as
\begin{eqnarray}
&&\ket{S(M+1,M)} \nonumber\\
%\ket{\frac{M+1}{2},\frac{M+1}{2}-1}=
&=&\frac{1}{\sqrt{M+1}}
\left\{
\ket{1}\ket{\frac{M}{2},\frac{M}{2}}+\sqrt{M}\ket{0} \ket{\frac{M}{2},\frac{M}{2}-1}
\right \}. \nonumber \\
\end{eqnarray}
Detailed investigation shows that $\ket{S(M+1, M)}$ is unavailable under $XX$ model without external magnetic fields. However, if the amount of evolution
time is set $t=\arctan \sqrt{M}/\sqrt{M}$, the outcome state is equivalent to $M+1$-particle $W$ state under a local unitary
transformation upon the center particle. To achieve the exact $W$ state, we hereby provide some alternative proposals with other
Hamiltonian models.
For $XXZ$ model without external applied magnetic field, if $\lambda=\frac{2}{1-M}$,  $t=\frac{\pi}{2\sqrt{M+1}}$, one could obtain
the state $\ket{S(M+1,M)}$ after the evolution. If the Hamiltonian is restricted in the $XX$ model, by adding a magnetic field of
$B=-2$, one could also approach the above state after a period of $t=\frac{\pi}{2\sqrt{M+1}}$.

A large group of physical systems including Josephson junction arrays, quantum dots and optical lattice
\cite {Experiment1, Experiment2, Duan, Probe} can be well described by the Heisenberg Hamiltonian,
therefore they are potential candidates for the implementation of our scheme. Without elaborate
sequence of time dependent fields, our scheme provides a more appealing method to generate arbitrary symmetric states in those experimental
systems where interactions are difficult to tune. Also, because all computation process is restricted in an isolated physical system, our
scheme further leads to a quantum computation protocol against decoherence.

In summary, we have presented a probabilistic scheme to generate arbitrary symmetric states based on spin networks involving single
particle measurements. Within the present scheme, no modulated pulses are required in the whole process. After the initial state,
a specific product state, is prepared upon a properly designed spin network, the whole system is supposed to undergo a period of free
dynamic evolution. By performing single particle measurements on supplementary particles, the required symmetric state is generated on
the target particles with a certain probability. More interestingly, we initiate a ``repeat-until-success" strategy with which one could
keep measuring the supplementary system until the required state is created. As a special case, we further demonstrated that
$M$ particle $W$ state can be produced with unit probability under various Hamiltonian models.

This work is supported by the National fundamental Research
Program (Grant No. 2001CB309300), the National
Science Fund for Distinguished Young Scholars (Grant No.
10425524). We also acknowledge the support by the European
Commission under Contact No. 007065 (Marie Curie Fellowship).

\end{document}